\documentclass[a4paper,12pt]{article}
\usepackage{epsfig}
\usepackage{amssymb}
\usepackage{amsfonts}
\usepackage{amsmath}
\usepackage{euscript}
\usepackage{verbatim}
\usepackage{latexsym}
\usepackage{graphicx}

	\usepackage{graphicx}	
	\usepackage{subfigure}
	\usepackage{wrapfig}
	\usepackage[T1]{fontenc}
	\usepackage{mathtools}
	\usepackage{float}

\usepackage{tikz}
\usetikzlibrary{shapes.geometric, arrows,patterns,snakes}
\tikzstyle{ellip} = [ellipse, minimum width=3cm, minimum height=1cm,text centered, draw=black]

\newskip\humongous \humongous=0pt plus 1000pt minus 1000pt

\newif\ifdtup

\catcode`\@=11

\@addtoreset{equation}{section}

\def\@normalsize{\@setsize\normalsize{15pt}\xiipt\@xiipt
\abovedisplayskip 14pt plus3pt minus3pt%
\belowdisplayskip \abovedisplayskip
\abovedisplayshortskip \z@ plus3pt%
\belowdisplayshortskip 7pt plus3.5pt minus0pt}

\def\small{\@setsize\small{13.6pt}\xipt\@xipt
\abovedisplayskip 13pt plus3pt minus3pt%
\belowdisplayskip \abovedisplayskip
\abovedisplayshortskip \z@ plus3pt%
\belowdisplayshortskip 7pt plus3.5pt minus0pt
\def\@listi{\parsep 4.5pt plus 2pt minus 1pt
     \itemsep \parsep
     \topsep 9pt plus 3pt minus 3pt}}

\relax

\catcode`@=12

\catcode`\@=11

\def\section{\@startsection{section}{1}{\z@}{3.5ex plus 1ex minus
   .2ex}{2.3ex plus .2ex}{\large\bf}}

\def\SymBoxes#1#2#3#4{\newdimen\un@t \un@t#3%
\raisebox{#1}{\rule{#2\un@t}{#4}\hskip-#2\un@t
\@tempdimb\un@t \advance\@tempdimb by-#4\@tempcntb#2\relax%
\@whilenum{\@tempcntb>0}\do{
\rule{#4}{\un@t}\hskip\@tempdimb \advance\@tempcntb by\m@ne}%
\hskip-#2\un@t \rule[\un@t]{#2\un@t}{#4}%
\rule[\un@t]{#4}{#4}\hskip-#4
\rule{#4}{\un@t}}\hskip-#4}                

\begin{document}

\newcommand{\beq}{\begin{equation}}
\newcommand{\eeq}{\end{equation}}
\newcommand{\bea}{\begin{eqnarray}}
\newcommand{\eea}{\end{eqnarray}}
\newcommand{\beas}{\begin{eqnarray*}}
\newcommand{\eeas}{\end{eqnarray*}}
\newcommand{\defi}{\stackrel{\rm def}{=}}
\newcommand{\non}{\nonumber}
\newcommand{\bquo}{\begin{quote}}
\newcommand{\enqu}{\end{quote}}
\renewcommand{\(}{\begin{equation}}
\renewcommand{\)}{\end{equation}}
\def \eqn#1#2{\begin{equation}#2\label{#1}\end{equation}}

\def\Fr{\left(1-\frac{2M}{r}\right)}
\def\Frv{\left(1-\frac{2M(v)}{r}\right)}
\def\e{\epsilon}
\def\IZ{{\mathbb Z}}
\def\IR{{\mathbb R}}
\def\IC{{\mathbb C}}
\def\IQ{{\mathbb Q}}
\def\de{\partial}
\def\Tr{ \hbox{\rm Tr}}
\def\H{ \hbox{\rm H}}
\def\HE{ \hbox{$\rm H^{even}$}}
\def\HO{ \hbox{$\rm H^{odd}$}}
\def\K{ \hbox{\rm K}}
\def\Im{ \hbox{\rm Im}}
\def\Ker{ \hbox{\rm Ker}}
\def\const{\hbox {\rm const.}}
\def\o{\over}
\def\im{\hbox{\rm Im}}
\def\re{\hbox{\rm Re}}
\def\bra{\langle}\def\ket{\rangle}
\def\Arg{\hbox {\rm Arg}}
\def\Re{\hbox {\rm Re}}
\def\Im{\hbox {\rm Im}}
\def\exo{\hbox {\rm exp}}
\def\diag{\hbox{\rm diag}}
\def\longvert{{\rule[-2mm]{0.1mm}{7mm}}\,}
\def\a{\alpha}
\def\dag{{}^{\dagger}}
\def\tq{{\widetilde q}}
\def\p{{}^{\prime}}
\def\W{W}
\def\N{{\cal N}}
\def\calB{\mathcal{B}}
\def\hsp{,\hspace{.7cm}}

\def\br{\nonumber\\}
\def\IZ{{\mathbb Z}}
\def\IR{{\mathbb R}}
\def\IC{{\mathbb C}}
\def\IQ{{\mathbb Q}}
\def\IP{{\mathbb P}}
\def \eqn#1#2{\begin{equation}#2\label{#1}\end{equation}}

\newcommand{\M}{\ensuremath{\mathcal{M}}
                    }
\newcommand{\oc}{\ensuremath{\overline{c}}}
\begin{titlepage}
\begin{flushright}
CHEP XXXXX
\end{flushright}
\bigskip
\def\thefootnote{\fnsymbol{footnote}}

\begin{center}
{\Large
{\bf An Alternate Path Integral for Quantum Gravity \\ 
\vspace{0.1in} 
}
}
\end{center}

\bigskip
\begin{center}
{
Chethan KRISHNAN$^a$\footnote{\texttt{chethan.krishnan@gmail.com}}, K. V. Pavan KUMAR$^a$\footnote{\texttt{kumar.pavan56@gmail.com}}, Avinash RAJU$^a$\footnote{\texttt{avinashraju777@gmail.com}}}
\vspace{0.1in}

\end{center}

\renewcommand{\thefootnote}{\arabic{footnote}}

\begin{center}

$^a$ {Center for High Energy Physics,\\
Indian Institute of Science, Bangalore 560012, India}\\

\end{center}

\noindent
\begin{center} {\bf Abstract} \end{center}
We define a (semi-classical) path integral for gravity with Neumann boundary conditions in $D$ dimensions, and show how to relate this new partition function to the usual picture of Euclidean quantum gravity. We also write down the action in ADM Hamiltonian formulation and use it to reproduce the entropy of black holes and cosmological horizons. A comparison between the (background-subtracted) covariant and Hamiltonian ways of semi-classically evaluating this path integral in flat space reproduces the generalized Smarr formula and the first law. This ``Neumann ensemble'' perspective on gravitational thermodynamics is parallel to the canonical (Dirichlet) ensemble of Gibbons-Hawking and the microcanonical approach of Brown-York. 



\vspace{1.6 cm}
\vfill

\end{titlepage}

\setcounter{footnote}{0}


\section{Introduction} 

A full definition of quantum gravity by starting with a Euclidean path integral for Einstein gravity is beset with various problems \cite{Carlip,CK}. This is possibly not surprising because the path integral over metrics is unlikely to be a full description of gravity if its UV completion is something like string theory. Instead of worrying about the subtleties in defining the path integral, what one tries instead is to interpret mostly only the physics around its saddles. In other words, most of the physics of Euclidean quantum gravity is obtained semi-classically. This semi-classical definition turns out to be quite rich and it has lead to many insights about black hole thermodynamics \cite{GHY} as well as holography \cite{Witten}.

This standard semi-classical definition of the gravitational path integral assumes that the metric that is being integrated over satisfies a Dirichlet boundary condition at the boundary of the spacetime manifold\footnote{There are further problems associated to infrared divergences that arise in flat space, but we will ignore such issues when they are not important, or overcome them via  suitable background subtraction. Or one can imagine working in asymptotically AdS spaces where the AdS length scale acts as an infrared regulator. In this case, we have to add further counter-terms. Some of this will be discussed in a companion paper \cite{AdSNeumann}.}. To make the variational principle well-defined and to obtain well-defined saddles with these boundary conditions, one has to then add a boundary term to the Einstein action, and this is the Gibbons-Hawking-York (GHY) boundary term \cite{GHY,York}. In this paper we will be interested in alternate definitions for the boundary conditions of the gravitational path integral, with our specific focus being the recently introduced Neumann gravity \cite{KR}. 


We will discuss the Neumann boundary condition in detail, for the gravity path integral. The saddles are well-defined under a variational principle that holds the canonical conjugate of the boundary metric (namely the boundary\footnote{sometimes also referred to as the quasi-local.} stress tensor density) fixed. An alternate to the GHY boundary term that can make such a Neumann action well-defined was introduced in \cite{KR}, and we will use that as our main tool. We will also work out a Hamiltonian formulation for such a theory and use it to compute the entropy of black holes and cosmological horizons. For this, we will put the discussion of boundary terms in a somewhat unified footing. Changing the boundary term corresponds to changing the ensemble, and we will see how the generalized Smarr formula and  first law can be seen using a background subtracted version of our Neumann path integral\footnote{We will only be working with flat space in the this paper.} and comparing the covariant and canonical results. We will also discuss the canonical ensemble of Gibbons-Hawking and the microcanonical path integral of Brown and York \cite{brown,quasilocal}, to emphasize the relation between various boundary terms.   

We will work with pure gravity 
but it seems evident that our approach should generalize straightforwardly when usual matter (\textit{i.e.}, scalar or gauge) fields are added, with their own boundary conditions. In an accompanying paper, we will discuss the Neumann path integral in AdS spaces \cite{AdSNeumann}. 

\section{Path Integral for Neumann Gravity}

Our starting point is the schematically defined path integral for the gravitational action with Neumann boundary conditions
\bea\label{density_of_states}
Z_N=\int D g\ e^{-S_N[g]}
\eea
where $S_N$ is the Einstein-Hilbert action with an appropriately constructed boundary term so that the saddles are well-defined with Neumann boundary conditions. We will first review a derivation of this action elaborating \cite{KR}. See also \cite{grumiller0, padmanabhan, Vassilevich, Park, McNees, HairyBH,marolf} for discussions on various boundary-related aspects of relevance to our problem. 

\subsection{Particle Mechanics}

We will start with the simple case of particle mechanics, and use the intuition there to generalize to our gravitational problem. We start with The Original Action:
\bea
S^p_D=\int^T dt \ \Big(\frac{1}{2}{\dot q}^2 -V(q) \Big)
\eea
We only show the upper boundary of the integral because we need to keep track of only one boundary for our purposes. The superscript $p$ stands for particle, and the subscript $D$ stands for Dirichlet. It is well-known to kindergarten-ers that this action is well-defined as a ``Dirichlet" problem\footnote{We emphasize however that by Dirichlet we only mean that we are holding $q$ fixed at the boundary, not $\dot q$. The fact that integration is along the time direction, and therefore this is not truly a boundary value problem is unimportant for the discussion at hand.}, namely when we hold $q$ fixed at $T$, we have a well-defined variational principle and we end up finding Newton's laws. To belabour the well-known, we find
\bea
\delta S^p_D&=&\int^T dt\left( -\Big(\ddot{q} +V'(q) \Big)\delta q +\frac{d}{dt}\Big(\dot{q} \delta q\Big)\right) \\
 &=&\int^T dt \Big(-\ddot{q} -V'(q) \Big)\delta q +(\dot{q} \delta q)|_T \label{Dvar}
\eea
Once we impose the ``bulk" equations of motion, we immediately see that setting $q=$ {\em any fixed quantity}, leads to a well-defined variational problem. This is the familiar ``Dirichlet" problem in particle mechanics, which is the source of all physics.

But even though not encountered in textbooks, we could also consider a ``Neumann" version of particle mechanics: where instead of holding the position, $q$, fixed at $T$ we wish to hold the velocity, $\dot {q}$, fixed. To accomplish this, we can add a ``boundary" term to the original action $S^p_D$  to get a Neumann action
\bea
S^p_N \equiv S^p_D - (q\ \dot{q})|_T
\eea
so that the resulting variation is
\bea
\delta S^p_N&=&\int^T dt \Big(-\ddot{q} -V'(q) \Big)\delta q -(q \delta \dot{q})|_T
\eea
and so that now, setting $\dot{q}=$ {\em any fixed quantity} leads to a well-defined variational principle and equations of motion.

In a somewhat more Hamiltonian notation, what we are really doing is to note that the boundary term in (\ref{Dvar}) is nothing but 
\bea
(\dot{q} \delta q)|_T=\pi_T \delta q_T, \ {\rm where} \ \pi_T(q_T,\dot{q_T}) \equiv \frac{\delta S^p_D}{\delta q_T}
\eea
where $q_T$ is the ``boundary" position variable, and then to define the Neumann action as 
\bea
S^p_N = S^p_D - \pi_T(q_T,\dot{q_T}) \ q_T.
\eea
Note that this is a boundary Legendre transform. In the particle mechanics case, this way of re-expressing the boundary velocity in terms of the boundary momentum might seem like introduction  of perhaps unnecessary formalism. Both correspond to fixing exactly the same physical quantity, after all. But it turns out that when we move on to gravity, this latter approach to the Neumann problem is the one that allows a natural generalization. To the best of our knowledge, a fully satisfactory Neumann boundary problem for standard Einstein-Hilbert gravity where one holds the {\em normal metric derivatives} fixed at the boundary is not known\footnote{Sometimes, setting the extrinsic curvature or its relatives to zero (eg., \cite{Zanelli,Takayanagi})is called a Neumann boundary condition. But it should be emphasized that this is a differential {\em constraint} on the boundary surface, and arbitrary boundaries will not allow it. An honest to God Neumann boundary condition should allow the metric derivatives to be {\em arbitrary} but {\em fixed} at the boundary. This is again perfectly analogous to particle mechanics, where we want the boundary position/velocity to be fixed, but not necessarily to some specific {\em value}.}. This is our main motivation for considering this formulation of Neumann, and it leads to a very general alternative to the GHY boundary term.

\subsection{Dirichlet Problem in Gravity}
Now we move on to our real interest, namely gravity. Our starting point is the standard Einstein-Hilbert action in D-dimensions which is given by\footnote{Our notations and conventions are that of \cite{poisson}}

\begin{equation}\label{eh_action}
S_{EH} = \frac{1}{2\kappa}\int_{\M}d^{D}x \sqrt{-g}(R-2\Lambda)
\end{equation}
where $\kappa=8\pi G$. Variation of Einstein-Hilbert action yields

\begin{equation}
\delta S_{EH} = \frac{1}{2\kappa}\int_{\M}d^{D}x \sqrt{-g}(G_{ab}+\Lambda g_{ab})\delta g^{ab} -\frac{1}{\kappa}\int_{\partial \M}d^{D-1}y \sqrt{|\gamma|} \varepsilon \left( \delta \Theta + \frac{1}{2}\Theta^{ij}\delta \gamma_{ij} \right)
\end{equation}
where $G_{ab} = R_{ab}-\frac{1}{2}R g_{ab}$ is the Einstein tensor, $\gamma_{ij} = g_{ab}e^{a}_{i}e^{b}_{j}$ is the induced metric on the boundary $\partial \M$ and $e^{a}_{i} = \frac{\partial x^a}{\partial y^i}$ is the coordinate transformation relating the boundary coordinates $y^i$ to the bulk coordinates $x^a$, and $\Theta = \gamma^{ij}\Theta_{ij}$ is the trace of the extrinsic curvature. $\varepsilon$ distinguishes the space-like and time-like hypersurfaces  and takes values $\varepsilon = \pm1$ for time-like and space-like boundaries respectively. We also assume that the boundaries are not null. The extrinsic curvature is defined as

\begin{equation}
\Theta_{ij} = \frac{1}{2}(\nabla_a n_b + \nabla_b n_a) e^{a}_{i}e^{b}_{j}
\end{equation}
 where $n_a$ is the unit normal to the boundary. The variational principle is spoiled by the offending surface term which does not vanish for a fixed boundary metric $\gamma_{ij}$. Therefore we need to add a boundary term to \eqref{eh_action} to make the variational principle well defined. This boundary piece is the Gibbons-Hawking-York term

\begin{equation}
S_{GHY} = \frac{1}{\kappa}\int_{\partial \M}d^{D-1}y \sqrt{|\gamma|} \varepsilon \Theta
\end{equation}
whose variation is given by 

\begin{equation}
\delta S_{GHY} = \frac{1}{\kappa}\int_{\partial \M}d^{D-1}y \sqrt{|\gamma|}\varepsilon \left(\delta \Theta + \frac{1}{2}\Theta \gamma^{ij}\delta \gamma_{ij} \right)
\end{equation}
Therefore the variation of total gravitational action yields

\begin{eqnarray}\label{dirichlet_variation}
\delta S_{D} &=& \delta S_{EH} + \delta S_{GHY} = \frac{1}{2\kappa}\int_{\M}d^{D}x \sqrt{-g}(G_{ab}+\Lambda g_{ab})\delta g^{ab} \\ \nonumber &-& \frac{1}{2\kappa}\int_{\partial \M}d^{D-1}y \sqrt{|\gamma|} \varepsilon  \left( \Theta^{ij} - \Theta \gamma^{ij}  \right)\delta \gamma_{ij}
\end{eqnarray}
Thus we find that the action $S_{D}$ is stationary under arbitrary variations of the metric in the bulk provided we satisfy the bulk equations of motion and the variations vanish on the boundary. 

For use in the next subsection, we define the canonical conjugate of the boundary metric as,
\begin{equation}
\pi^{ij} \equiv \frac{\delta S_{D}}{\delta \gamma_{ij}} = -\frac{\sqrt{|\gamma|}}{2\kappa} \varepsilon (\Theta^{ij}-\Theta \gamma^{ij})
\end{equation}
The variation of \eqref{dirichlet_variation} can thus be expressed as

\begin{equation}\label{variation}
\delta S_{D} = \frac{1}{2\kappa}\int_{\M}d^{D}x \sqrt{-g}(G_{ab}+\Lambda g_{ab})\delta g^{ab} + \int_{\partial \M}d^{D-1}y\; \pi^{ij} \delta \gamma_{ij}
\end{equation}

\subsection{Neumann Problem in Gravity}

The aim of the previous section was to express the variation of Dirichlet action in a suggestive form which is suitable for moving to Neumann problem. We take the view that a well-defined Neumann problem is one where instead of holding the boundary metric, its canonical conjugate is held fixed (while satisfying the bulk equations of motion). This can be accomplished by adding yet another term to the Dirichlet action. The form of the new term is suggested by \eqref{variation} as follows

\begin{equation}
S_{N} = S_{EH} + S_{GHY} -\int_{\partial \M} d^{D-1}y\ \pi^{ij}\gamma_{ij}
\end{equation}
The variation of $S_{N}$ yields,

\begin{equation}
\delta S_{N} = \frac{1}{2\kappa}\int_{\M}d^{D}x \sqrt{-g}(G_{ab}+\Lambda g_{ab})\delta g^{ab} - \int_{\partial \M}d^{D-1}y\ \delta \pi^{ij} \gamma_{ij}
\end{equation}
In terms of the extrinsic curvature, the action takes the form

\begin{equation}\label{neumann_action}
S_{N} = \frac{1}{2\kappa}\int_{\M}d^{D}x \sqrt{-g}(R-2\Lambda) + \frac{(4-D)}{2\kappa}\int_{\partial \M}d^{D-1}y \sqrt{|\gamma|}\varepsilon \Theta
\end{equation}

This is the Neumann action that we will use to give a semi-classical definition to the Neumann path integral for gravity. 

\subsection{Sewing Together Path Integrals}

Before moving on to other things, we make one comment about how one can sew path integrals together to build a new path integral in this picture. The Dirichlet path integral for gravity enjoys the sewing property, wherein if we cut the spacetime manifold $\cal{M}$ into two pieces ${\cal M}_1$ and ${\cal M}_2$ joined at a hypersurface $\Sigma$, then the total path integral can be viewed as being ``sewed" together from the two separate path integrals, 
\bea
Z_D^{\cal M}=\int [d g_0] Z_D^{{\cal M}_1}[g_0] Z_D^{{\cal M}_2}[g_0].
\eea
The crucial fact here is that the extrinsic curvature has a sign that is controlled by the normal to the surface $\Sigma$, it occurs with compensating sign on the two $Z$ pieces on the right hand side. It is for the same reason, similar construction holds for Neumann theory also and the sewing property is satisfied. 
\bea
Z_N^{\cal M}=\int [d \pi_0] Z_N^{{\cal M}_1}[\pi_0] Z_N^{{\cal M}_2}[\pi_0].
\eea
where the canonical conjugate is held fixed at $\Sigma$. It will be interesting to evaluate this path integral explicitly in detail (perhaps in 2+1 dimensions where the Neumann action translates into a pure Chern-Simons theory \cite{KR}) to see whether the boundary conditions force the presence of our boundary term. Related computations in lower dimensions and in other contexts have been done in \cite{CarlipCMP, WittenFactor, Willy}.

\section{Hamiltonian Formulation of Neumann Gravity}\label{ADM_sec}

The Arnowitt-Deser-Misner (ADM) approach is a space+time split of the field variables in gravity that is useful as a natural starting point for the Hamiltonian formulation of general relativity. We wish to write down the Neumann action in this language, with applications in later sections in mind. 

We consider manifolds which are like box, \textit{i.e.} are cut-off at finite spatial distance so that the boundary is time-like, denoted $\mathcal{B}$. The spatial section of the boundary $\mathcal{B}$ is denoted $B$. The covariant action is given by
\begin{equation}\label{cov_action}
S_N = \frac{1}{2\kappa}\int_{\mathcal{M}}d^Dx\;\sqrt{-g}({}^{(D)}R - 2\Lambda) + \frac{(4-D)}{2\kappa}\int_{\mathcal{B}}d^{D-1}y\;\sqrt{-\gamma}\Theta
\end{equation}
where $\Theta$ is the extrinsic curvature of $\mathcal{B}$ and $\gamma_{ij}$ is the induced metric. ADM approach relies on foliating the D-dimensional spacetime $(\mathcal{M},g_{\alpha \beta})$ by $(D-1)$-dimensional spatial hypersurface $(\Sigma_t,h_{ab})$, labelled by the time parameter $t$. The time-like unit normal to the hypersurface $\Sigma_t$ is denoted $u^{\alpha}$ and satisfies, $u^{\alpha}u_{\alpha}=-1$. The spacetime metric can be expressed as
\begin{equation}\label{adm_metric_full}
ds^2 \equiv g_{\alpha \beta}dx^{\alpha} dx^{\beta} = -N^2 dt^2 + h_{ab}(dy^a + N^{a}dt)(dy^b + N^{b}dt) 
\end{equation}
where $N$ is the lapse function, $N^{a}$ is the shift vector and $h_{ab}$ is the induced metric on the hypersurface $\Sigma_t$. The induced metric $\gamma_{ij}$ can also be split as
\begin{equation}\label{gamma_ADM}
ds^2 \equiv \gamma_{ij}dx^{i} dx^{j} = -N^2 dt^2 + \sigma_{AB}(d\theta^A + N^{A}dt)(d\theta^{B} + N^{B}dt)
\end{equation}
where $\sigma_{AB}$ is the induced metric on $B$. The space-like boundary which is at the initial time $t_i$ and final time $t_f$ is ignored. The reason for this as follows. We can always choose to work with a box where the hypersurfaces $\Sigma_t$ and $\mathcal{B}$ are mutually orthogonal, \textit{i.e.} $u^{\alpha} r_{\alpha} = 0$, where $r^{\alpha}$ is the radial outward pointing unit vector. This allows us to split the boundary of the spacetime manifold into three parts $\partial \mathcal{M}=\mathcal{B} \cup \Sigma_{t_i}\cup \Sigma_{t_f}$. Since we are ultimately interested in horizon entropy calculations, where the (Wick rotated) time is periodically identified (as we will explain), the surface contributions coming from  $\Sigma_{t_{i}}$ and $\Sigma_{t_{f}}$ plays no role.
The Ricci scalar of the bulk spacetime can be decomposed as
\begin{equation}\label{ricci_decomp}
{}^{(D)} R = {}^{(D-1)}R + K^{ab}K_{ab} - K^2 - 2\nabla_{\alpha}\left( u^{\beta} \nabla_{\beta}u^{\alpha} - u^{\alpha} \nabla_{\beta} u^{\beta} \right)
\end{equation}
where $K_{ab}$ is the extrinsic curvature of $\Sigma_t$. Substituting the above expression into \eqref{cov_action}, we obtain \footnote{For an analogous discussion of Dirichlet problem in ADM decomposition, see \cite{poisson}}
\begin{eqnarray}\label{adm_split_action}
(2\kappa)S_{N} &=& \int_{\mathcal{M}}d^D x \sqrt{-g}\left[ {}^{(D-1)}R - 2\Lambda + K^{ab}K_{ab} - K^2 \right] \\ \nonumber &-& 2 \int_{\mathcal{B}} \left( u^{\beta} \nabla_{\beta}u^{\alpha} - u^{\alpha} \nabla_{\beta} u^{\beta} \right)d\Sigma_{\alpha} + (4-D)\int_{\mathcal{B}} d^{D-1}y \sqrt{-\gamma}\Theta
\end{eqnarray}

On the hypersurface $\mathcal{B}$, the measure $d\Sigma_a$ is given by $d\Sigma_a = r_a \sqrt{-\gamma}d^{D-1}y$.  The surface integral on $\mathcal{B}$ gives
\begin{equation}
-2\int_{\mathcal{B}} \left(u^{\beta} \nabla_{\beta}u^{\alpha} - u^{\alpha} \nabla_{\beta} u^{\beta} \right)d\Sigma_{\alpha} = 2\int_{\mathcal{B}} d^{D-1}y \sqrt{-\gamma} u^{\alpha} u^{\beta} \nabla_{\beta} r_{\alpha} 
\end{equation}
where we have used the fact that $u^{\alpha} r_{\alpha} = 0$. The action \eqref{adm_split_action} now becomes
\begin{eqnarray}\label{adm_split_action_1}
(2\kappa)S_{N} &=& \int_{\mathcal{M}}d^D x \sqrt{-g}\left[ {}^{(D-1)}R - 2\Lambda + K^{ab}K_{ab} - K^2 \right] \\ \nonumber &+& 2 \int_{\mathcal{B}} d^{D-1}y \sqrt{-\gamma}u^{\alpha} u^{\beta} \nabla_{\beta} r_{\alpha} + (4-D)\int_{\mathcal{B}} d^{D-1}y \sqrt{-\gamma}\Theta
\end{eqnarray}
The two surface terms in the above expression can be rearranged as follows. From the definition of extrinsic curvature \cite{poisson}, it follows that
\begin{eqnarray}
\Theta + u^{\alpha}u^{\beta} \nabla_{\beta} r_{\alpha} = \sigma^{AB}(\nabla_{\beta} r_{\alpha} e^{\alpha}_{A}e^{\beta}_{B}) = k_{AB}\sigma^{AB} \equiv k
\end{eqnarray}
where $\sigma_{AB}$ is the induced metric on the boundary $\partial \Sigma_t$ (see \eqref{gamma_ADM}), $k_{AB}=(\nabla_{\beta} r_{\alpha}) e^{\alpha}_{A}e^{\beta}_{B}$ is the extrinsic curvature of $\partial \Sigma_t$ embedded in $\Sigma_t$. $e^{\alpha}_{A}\equiv \frac{\partial x^{\alpha}}{\partial \theta^A}$ is the projector relating the bulk coordinates\footnote{See eqn. (\ref{adm_metric_full}) for definition of $x^\alpha$.} $x^{\alpha}$  to the $\partial \Sigma_t$ coordinates $\theta^A$.

Using the expression for determinants $\sqrt{-g}=N\sqrt{h}$ and $\sqrt{-\gamma}=N\sqrt{\sigma}$, \eqref{adm_split_action_1} can be expressed as 
\begin{eqnarray}\label{adm_split_action_2}
(2\kappa)S_{N} &=& \int^{t_f}_{t_i}dt \left[ \int_{\Sigma_t}d^{D-1} y\; N\sqrt{h}\left( {}^{(D-1)}R - 2\Lambda + K^{ab}K_{ab} - K^2 \right) \right. \\ \nonumber &+& \left. 2 \int_{B} d^{D-2}\theta \; N\sqrt{\sigma}k + (2-D)\int_{\partial \Sigma_t} d^{D-2}\theta \; N \sqrt{\sigma}\Theta \right]
\end{eqnarray}
To obtain the action in terms of canonical variables, we have to introduce the conjugate momentum for $h_{ab}$. This is given by
\begin{equation}
p^{ab} \equiv \frac{\partial}{\partial \dot{h}_{ab}}(\sqrt{-g}\mathcal{L}_G) = \frac{\sqrt{h}}{2\kappa}(K^{ab}-Kh^{ab})
\end{equation}

The extrinsic curvature $\Theta$ can be be split into two pieces as
\begin{equation}
\Theta = \Theta^{ij}\gamma_{ij} = \Theta^{ij}(\sigma_{ij} -u_i u_j\Theta^{ij}) = k-u_i u_j \Theta^{ij} = k+\frac{r^a\partial_a N}{N}
\end{equation}
With this expressions for $p^{ab}$ and $\Theta$, the action can be written as
\begin{eqnarray}\label{Neumann_ADM_1}
S_N &=& \int_{\mathcal{M}}d^D x \left( p^{ab}\dot{h}_{ab} - NH - N_a H^a \right) +  \int_{\mathcal{B}}d^{D-1}y \sqrt{\sigma}\left(\frac{N k}{\kappa} - \frac{2N^a r^b p_{ab}}{\sqrt{h}} \right) \\ \nonumber &+& \frac{(2-D)}{2\kappa}\int_{\mathcal{B}}d^{D-1}y\;N\sqrt{\sigma}(k+\frac{r^a\partial_a N}{N})
\end{eqnarray}
where $H$ and $H^a$ are the Hamiltonian and momentum constraints respectively, whose exact expressions are given by
\begin{eqnarray}
H &=& \frac{\sqrt{h}}{2\kappa}\left( K^{ab}K_{ab} - K^2 - {}^{(D-2)}R + 2\Lambda \right) \\ \nonumber
H^a &=& -\frac{\sqrt{h}}{\kappa}D_b(K^{ab}-Kh^{ab}) \\ \nonumber
\end{eqnarray}

The two boundary integrals can be now combined and expressed in terms of canonical variables as
\begin{eqnarray}\label{adm_action_neumann}
S_N &=& \int_{\mathcal{M}}d^D x \left( p^{ab}\dot{h}_{ab} - NH - N_a H^a \right) +  \int_{\mathcal{B}}d^{D-1}y \sqrt{\sigma}\left(\frac{N\varepsilon}{2}  - N^a j_a + \frac{N}{2}s^{ab}\sigma_{ab} \right)
\end{eqnarray}
where $\sqrt{\sigma}\varepsilon$, $\sqrt{\sigma}j_a$ and $N\sqrt{\sigma}s^{ab}/2$ are the momenta conjugate to $N$, $N^{a}$ and $\sigma_{ab}$, respective \cite{brown}. They are defined as
\begin{eqnarray}\label{canonical_variables}
\varepsilon &=& \frac{k}{\kappa}, \quad  j_a = \frac{2}{\sqrt{h}}r_b p_{\;a}^{b} \\ \nonumber s^{ab} &=& \frac{1}{\kappa}\left[k^{ab} - \left(\frac{r^a \partial_a N}{N} + k\right)\sigma^{ab} \right]
\end{eqnarray}

\section{Boundary Terms, Ensembles and Partition Functions}

We will view consistent boundary terms for Einstein-Hilbert action as definitions of thermodynamic ensembles arising in Euclidean quantum gravity. In this spirit, we should be able to reproduce the thermodynamics of horizons with our Neumann path integral/partition function. We will see that this is indeed the case. To give context we will discuss the canonical ensemble of Gibbons-Hawking and the microcanonical path integral of Brown-York\footnote{In what follows we will refer to the latter as the Brown-York path integral, because the term "microcanonical" can give rise to confusion in an AdS/CFT context: fixing the total CFT energy is different from fixing the energy density at the boundary of AdS.}. The philosophy of the latter is essential for our purposes and since this is not widely known we will review it here. Our discussion of the Gibbons-Hawking ensemble will not follow \cite{GHY}. Instead we will view it as a Laplace transform from the Brown-York ensemble \cite{brown}. We will review that as well.

It is convenient to start with the Brown-York path integral \cite{brown} whose notations we largely follow, except for the sign conventions for extrinsic curvature, for which we follow \cite{poisson}.


\subsection{Brown-York Ensemble}

We will first show how the Brown-York path integral for the gravitational field can be used to compute the entropy, and then in the next subsection relate it to the (grand) canonical approach of Gibbons-Hawking. After that, we will also clarify how our Neumann path integral naturally fits into all of this. 

We will work with the ADM formulation, as in the previous section to discuss the various ensembles. In this language the microcanonical action of Brown-York takes the form \cite{brown}
\bea \label{Brown-York-action}
S_{BY} = \int_{\mathcal{M}}d^D x\;\left( p^{ab}\dot{h}_{ab} - NH - N_a H^a \right) 
\eea
This is just the bulk piece of the ADM action we wrote down before. The claim is that this action (ie., with no ADM boundary terms) can be used to compute a (microcanonical) density of states via
\bea
\nu[\epsilon,j,\sigma]=\sum_M\int D[H]\  \exp(i S_m) \label{mupath}
\eea
Here, the density of states is a functional of the boundary data
\bea
(\epsilon, \ \ j_a, \ \ \sigma_{ab}) \label{micro-indep}
\eea
that were defined in the previous section. In other words, the action (\ref{Brown-York-action}) is chosen because its variations vanish when these quantities are held fixed at the boundary.

The integration measure is over all field configurations, see \cite{brown} for details of the  notation in this functional integral: only the general idea is important to us here. A crucial point in this construction is that the time direction needs to be periodically identified, but the periodicity is arbitrary. We will not discuss the origin of this construction in detail even though it is straightforward. But we make one comment about the dependence on $\epsilon,j_a,\sigma_{ab}$: in non-gravitational systems, the Hamiltonian is the generator of time-translations. However for gravitating systems, the bulk Hamiltonian vanishes as a constraint. Therefore, it is natural to hold the energy surface density, momentum surface density as well as the boundary metric fixed when defining a version of the gravitational microcanonical action. This is indeed what (\ref{Brown-York-action}) is. A further comment is that the periodicity in time arises as a standard consequence of the trace involved in the definition of the quantum density of states, see eg. \cite{brown,Chaichian}.


Now we will use this path integral to explore the thermodynamics of black holes. We can work with the standard Schwarzschild metric to compute the entropy, as is usually done to illustrate the point, but we will instead keep the discussion general and work with more general stationary rotating black holes as done by \cite{brown}. This approach naturally generalizes to computation of entropy of cosmological horizons like de Sitter space as well, as we will illustrate later.

The Lorentzian black-hole metrics have the boundary topology $\mathcal{B}=S^{(D-2)}\times I$, but to connect with the above-mentioned periodicity in time means that we need to consider a manifold with boundary topology $S^{(D-2)}\times S^{1}$. Following \cite{brown}, there is a related "complexified" Lorentzian black hole metric with periodically identified time 
which can be used for entropy calculations. Even though the metric is not "real", the motivation for introducing it is that it can be used for steepest descents approximation to the functional integral by distorting the $N$ and $N^a$ contours in the complex plane.   

A general Lorentzian black hole metric is of the form
\begin{equation}\label{lor_bh}
ds^2 = -N^2 dt^2 + h_{ab}(dx^a +N^a dt)(dx^b +N^b dt)
\end{equation}
and the accompanying complex metric is given by
\begin{equation}\label{com_bh}
ds^2 = -(-i\tilde{N})^2 dt^2 + \tilde{h}_{ab}(dx^a - i\tilde{N}^a dt)(dx^b - i\tilde{N}^b dt)
\end{equation}
Thus the complex metric is related to the Lorentzian metric with the identification $N=-i\tilde{N}$ and $N^a = -i\tilde{N}^a$. We assume that the black hole metric is stationary so that $\tilde{N}$ and $\tilde{N}^a$ are independent of time. Since Einstein's equations are analytic differential equations, if \eqref{lor_bh} is a solution, then so is \eqref{com_bh}. The locus of points $\tilde{N}=0$ describes a hypersurface called "bolt" and the $t={\text constant}$ foliations degenerate at that point\footnote{For Schwarzschild, this just becomes the usual cigar geometry where one demands regularity at the tip.}. If we choose "corotating" spatial coordinates then $\tilde{N}_a$ vanish on the horizon and thus near the bolt, the metric takes the form
\begin{equation}
ds^2 \approx \tilde{N}^2 dt^2 + \tilde{h}_{ab}dx^a dx^b \label{corotframe}
\end{equation}
and demanding the absence of conical singularities at the bolt gives the relation
\begin{equation}\label{perodicity_condition}
r^a \partial_a \tilde{N} = \frac{2\pi}{P}
\end{equation}
where $P$ is the periodicity of the time coordinate. The $t={\text constant}$ hypersurface has the topology $\Sigma_t = I\times S^{(D-2)}$ and the singularity can be interpreted as a puncture on $\Sigma_t$. The bolt seals the spatial manifold but it needs an appropriate boundary condition to be specified there. Note also that it is in the precise determination of the temperature via Euclidean regularity that all the ensembles get a {\em semi}-classical interpretation. One can obtain versions of the first law in all ensembles (including Neumann, as we will show later) purely classically, but the entropy and temperature always appear multiplied together. To fix their relative numerical factor one needs the semi-classical input\footnote{We thank Ghanashyam Date for emphasizing this to us.}.

So we start with the action \eqref{neumann_action} and follow the step of canonical decomposition of previous section, but with an inner boundary which is at the horizon as specified by \cite{brown,martinez}: they demand that no extra boundary piece be added at the horizon other than what already arises from the Einstein-Hilbert term in the Hamiltonian form. They argue that this is natural because in the Lorentzian section the horizon is smooth: there exists a free-fall coordinate system (Kruskal, say) where the principle of equivalence explicitly holds and there is no (coordinate) singularity. We will find that this boundary condition indeed yields the right black hole thermodynamics in all the ensembles, including Neumann. In any event, this yields the following Brown-York action for the black hole geometry:
\begin{eqnarray}
S_{BY}&=& \int_{\mathcal{M}}d^D x\;\left( p^{ab}\dot{h}_{ab} - NH - N_a H^a \right) +S_{horizon},
\end{eqnarray}
where $S_{horizon}$ is given by \cite{brown,martinez}
\bea
S_{horizon}\equiv \int_{\mathcal{H}}d^{D-1}y\;\sqrt{\sigma}\left(\frac{r^a \partial_a N}{\kappa} + \frac{2r_a N_b p^{ab}}{\sqrt{h}} \right)\label{BH_dos_action}
\eea
This action, just as the (\ref{Brown-York-action}), has the property that the boundary  data at $\mathcal{B}$ is specified by fixed $\varepsilon$, $j_a$ and $\sigma_{ab}$. In this section, we will assume that the boundary metric $\sigma_{ab}$ is axi-symmetric. 

Now we are ready to evaluate the action integral on the solution. Notice that the bulk integral is identically zero by virtue of being a stationary metric and the fact that it satisfies the Hamiltonian and momentum constraints. Thus the Brown-York action for the black hole yields
\begin{equation}
S^{BH}_{BY}
= -\frac{i}{\kappa}\int_{0}^{P}dt\int d^{d-2}\theta\;\sqrt{\sigma} r^{a}\partial_a \tilde{N}
\end{equation}
Since we are working in the co-rotating frame (\ref{corotframe}), $N^a$ at the horizon is zero, the second term in (\ref{BH_dos_action}) doesn't contribute. Therefore this discussion that reproduces the entropy as in \cite{brown} is not really a strong check of the second piece in (\ref{BH_dos_action}). However, we will see that when we move on to other ensembles (both Gibbons-Hawking and Neumann), this term will precisely reproduce the thermodynamics of black holes: one can view this as a further piece of evidence that the horizon boundary condition suggested in \cite{brown, martinez} is a reasonable one. We will see that in the co-rotating frame, the contributions of some of the thermodynamic quantities arise solely from the boundary, whereas in other frames it is shared between the boundary and the horizon. In both cases, it happens in such a way what the appropriate generalized Smarr formula holds.

Upon using the periodicity condition \eqref{perodicity_condition} the above expression for the entropy leads to
\begin{equation}
S^{BH}_{BY}
 = -\frac{2\pi i}{\kappa}A_H = -\frac{i}{4}A_H
\end{equation}
The advantage of this form is that since the functional integral (\ref{mupath}) that is defined by $S_{BY}$ has the interpretation as a microcanonical density of states. Therefore, from the standard ideas of statistical mechanics, a saddle point argument yields the entropy of the system to be
\begin{equation}
\mathcal{S}[\varepsilon,j_a,\sigma_{ab}] \approx iS_{BY}
\end{equation}
This gives the requisite relation between the entropy and the area of the horizon:
\begin{equation}
\mathcal{S}[\varepsilon,j_a,\sigma_{ab}] \approx iS_{BY}^{BH} = \frac{A_H}{4}
\end{equation}
This is the Brown-York derivation of the microcanonical entropy, presented here with an eye towards generalization to other ensembles, especially Neumann.

We make one comment before moving on to other boundary term choices. The time periodicity in the Brown-York approach arose because the microcanonical density of states involves a trace over states. But instead of putting this periodicity in from the start, one can also consider the Brown-York path integral (in Euclidean, or more precisely complexified, space) without the periodicity as the starting point. If one demands that the relevant saddles of this path integral are smooth spacetimes, then one again gets the time-periodicity from the smoothness of the ``bolt'' as discussed above. This is the approach we will adopt for other ensembles. The saddles of the Euclidean path integrals will always fix a periodicity in imaginary time, irrespective of the boundary terms involved: this is because the smoothness at the horizon is what fixes it.


\subsection{Gibbons-Hawking Ensemble}

Einstein Hilbert action with the GHY boundary term (aka the Dirichlet aka grand canonical action) that we discussed earlier can be computed in the ADM formulation (analogous to what we did for Neumann in the previous section) to be\footnote{We give it the subscript $g$ for grand canonical instead of $D$ for Dirichlet here to emphasize its role in thermodynamics.}
\bea
S_g = \int_{\mathcal{M}}d^D x\;\left( p^{ab}\dot{h}_{ab} - NH - N_a H^a \right) + \int_{\mathcal{B}} d^{D-1}y \sqrt{\sigma}(N \epsilon- N^a j_a) \label{GHYADM}
\eea
In the (grand) canonical ensemble, the boundary data is specified by fixing the potentials at the boundary, for example, an axi-symmetric black hole is specified by its inverse temperature $\beta$ and (co-rotating) angular velocity $\Omega_H$. The various black hole quantities are then to be thought of as functions of these potentials.  One can think of this ensemble as one where
\bea
(N, \  N^a, \ \sigma_{ab}) \ {\rm or} \ {\rm equivalently} \ (\beta, \  \omega,  \sigma_{ab}) \label{canon-indep}
\eea
are held fixed; $\omega$ will be defined momentarily.

One can relate the Brown-York approach of the previous subsection to the (grand) canonical approach of Gibbons-Hawking as follows  (see \cite{brown}). We define $\phi^a$ to be the axial Killing vector\footnote{The existence of such a Killing vector is implicit in the Gibbons-Hawking paper, and our goal is to make connection with it.} for the boundary metric $\sigma_{ab}$. The momentum density along this Killing direction is then given by $\sqrt{\sigma} j_a \phi^a$. The (grand) canonical partition function can be obtained from the microcanonical partition function (\ref{mupath}) via 
\bea
Z_g[\beta, \beta\omega,\sigma]=\int D[\sqrt{\sigma}\epsilon]D[\sqrt{\sigma}j_a\phi^a]\ \nu[\epsilon,j,\sigma]\ \exp\left[\int d^{D-2}\theta \sqrt{\sigma} \beta(\varepsilon-\omega j_a\phi^a)\right] 
\eea
The steepest descents of $Z_g$ is then given by the simultaneous solutions of

\begin{equation}
\frac{\delta {\cal S}}{\delta(\sqrt{\sigma}\varepsilon)}=-\beta, \qquad \frac{\delta {\cal S}}{\delta(\sqrt{\sigma}j_a \phi^a)}=\beta\omega
\end{equation}
In the saddle point approximation, this yields
\bea
iS_g \approx \ln Z_g \approx {\cal S}-\int d^{D-2}\theta \sqrt{\sigma} \beta(\epsilon-\omega j_a\phi^a)
\eea
Using the (\ref{mupath}) one can write 
\bea
Z_g[\beta, \beta\omega,\sigma]=\sum_M\int D[H]\  \exp(i S_g)\label{grandcanpartfn}
\eea
where the $S_g$ is evaluated with 
\bea
\int dt N|_B=-i\beta, \ \
\int dt N^\phi|_B=-i\beta\omega, \label{betaomega}
\eea
where $N^\phi$ is the shift along the $\phi^a$ direction and $B$ is the spatial slice of ${\cal B}$ on which we have the metric $\sigma_{AB}$. So instead of $\sqrt{\sigma}\varepsilon$ and $\sqrt{\sigma}j_a$ (which are to be thought of as conjugate fields living in the phase space) being fixed at $B$, in the (grand) canonical picture we have their potentials fixed at $B$. 

Notice that for a stationary black holes we can assume that $\beta$ is a constant on the boundary\footnote{See however, Lewkowycz and Maldacena \cite{Maldacena} who investigate situations where this assumption does not hold. Indeed, the construction of this subsection as well as the entire spirit of this paper can be generalized to non-constant fields at the boundary. The Gibbons-Hawking (aka standard grand canonical) ensemble is a special case of the Lewkowycz-Maldacena ensemble.}. This is because stationary black holes have time-like Killing vector which gets identified as the generator of $U(1)$ isometry under the Wick rotation. Note also that as the boundary $B$ is taken to radial infinity, since $N \rightarrow 1$ for stationary flat space black holes, $\beta$ is fixed by the periodicity of the time circle. This means that since we fixed this periodicity in the Brown-York ensemble via the smoothness of the "bolt", before doing our Laplace transform, it will be the same as the $P$ that we discussed in the previous subsection. 

Similarly, $\omega$ is defined as the angular velocity measured by the so-called Zero Angular Momentum Observers (ZAMOs) \cite{brown}. Note however, that asymptotically $\omega$ goes to zero, for the Kerr black hole \cite{poisson}. For a rotating black hole, as we saw in the previous subsection when discussing the smoothness of the ``bolt'', as well as for dynamical and thermodynamical reasons \cite{HawkingHunter,CKTomogram}, it is more reasonable to think of the black hole in a box that is co-rotating-with-the-horizon\footnote{There are some problems here, related to the fact that a co-rotating frame becomes superluminal at a finite radius, and therefore it is not really possible to define a fully consistent thermodynamics for rotating black holes in flat space. A related observation is that there is no fully satisfactory Hartle-Hawking state for the flat space Kerr black hole \cite{KayWald,CK-BH-Rot}. But these problems are usually glossed over and a formal treatment of thermodynamics is unaffected by them. Our discussion should be taken in that spirit, we will see in a follow up paper \cite{AdSNeumann} that for AdS black holes these problems have natural solutions.}. In such a frame, the second equation above gets modified to
\bea
\int dt N^\phi|_B=i\beta(\Omega_H-\omega)\rightarrow i\beta \Omega_H, \label{OmegaH}
\eea
where the last step takes into account the fact that asymptotically $\omega$ is zero. $\Omega_H$ is the horizon angular velocity \cite{poisson}.

One can also obtain the grand canonical path integral directly by defining a Euclidean path integral with (\ref{GHYADM}) as the action. Since this is the approach we will take when defining the Neumann partition function\footnote{As we will explain in the next subsection, in terms of canonical variables, the Neumann boundary term is not a simple Legendre transform like it was in covariant variables. So the transformation relating the two ensembles is non-trivial, unlike in the transformation to Gibbons-Hawking from Brown-York. We will not pursue this further here.} let us emphasize it. Demanding that the (complex) saddles of the action are regular again fixes the periodicity of the time circle to be the inverse Hawking temperature. After complexification we get the barred versions of (\ref{betaomega}, \ref{OmegaH}) as definitions of $\beta$ and $\omega$: note that $N$ and $N^\phi$ are fixed at the boundary.


In the (grand) canonical ensemble then, it is natural to have the formula 
\bea 
\frac{-S_g}{\beta}=M- T{\cal S} -\Omega_H J,\label{GHFreeEnergy}
\eea 
which directly lead to useful relations like ${\cal S}=(\beta\partial_\beta-1) S_g$. We have defined $T=1/\beta$. 
Gibbons and Hawking motivate (\ref{GHFreeEnergy}) by noting that the free energy associated to a grand canonical partition function is {\em defined} by the right hand side of (\ref{GHFreeEnergy}). Then they make a saddle-point approximation to (\ref{grandcanpartfn}) to obtain the left hand side. At this stage, they note that covariantly evaluating the left hand side of (\ref{GHFreeEnergy}) on black hole solutions after background subtraction leads to the generalized Smarr formula. This is how \cite{GHY} relate the classical general relativity of black holes with black hole thermodynamics. 

But instead of taking the right hand side of (\ref{GHFreeEnergy}) as a definition of the thermodynamic potential like \cite{GHY} does, one can instead look at it as an explicit evaluation of the action in the {\em canonical} (aka ADM aka Hamiltonian) approach. We will see in detail that this reproduces the RHS of (\ref{GHFreeEnergy}), in the next section. This means that one can view the emergence of the generalized Smarr formula from (\ref{GHFreeEnergy}) as a result of comparing the covariant and Hamiltonian ways of (saddle point) evaluating the Gibbons-Hawking partition function (\ref{grandcanpartfn}).


The reason why we emphasize this perspective is that it is this approach that has a natural adaptation to the Neumann case as we will see in a later section.  We will evaluate background subtracted version of the Neumann action for the flat space black holes in various dimensions, and we will see that the generalized Smarr formula arises, parallel to the canonical ensemble result in \cite{GHY}.

\subsection{Neumann Ensemble}

The basic idea that we used in the previous section to obtain the Gibbons-Hawking path integral from the Brown-York path integral is the fact that in Statistical mechanics, given the microcanonical density of states, other thermodynamic potentials can be obtained by suitable functional Laplace transforms. However, the Neumann ensemble is best thought of as a mixed ensemble. We can see this as follows.

Eqn (\ref{adm_action_neumann}) is the Hamiltonian (ADM) version of the Neumanna action which we will find useful in what follows. First note that the variation of the action \eqref{adm_action_neumann} with respect to the canonical variables gives
\begin{eqnarray}
\delta S_N &=& ( \text{eq. of motion})\label{neu_act_var} \\ \nonumber &+& \int_{\cal B}d^{D-1}y \left[ \delta N (\sqrt{\sigma}\varepsilon/2) - N \delta(\sqrt{\sigma}\varepsilon/2) - \delta N^a (\sqrt{\sigma}j_a) + \delta(N\sqrt{\sigma}s^{ab}/2)\sigma_{ab} \right]
\end{eqnarray}
The first two terms can be combined into $\delta N (\sqrt{\sigma}\varepsilon/2) - N \delta(\sqrt{\sigma}\varepsilon/2)=N^2 \delta\left(\frac{\sqrt{\sigma}\varepsilon}{2N}\right)$. This means that one can view the Neumann action and the associated partition function loosely\footnote{Loosely, because we are not being careful about the boundary symplectic structure in making this variable redefinition. Our discussion will not rely on this subtlety.} as an ensemble in which $\frac{\sqrt{\sigma}\varepsilon}{N}, N^a$ and $N\sqrt{\sigma}s^{ab}$ are held fixed. This should be contrasted to the Brown-York and Gibbons-Hawking ensembles which are defined by (\ref{micro-indep}) and (\ref{canon-indep}) respectively, see \cite{brown}.

We noted before that in the covariant formalism, the Gibbons-Hawking action and Neumann actions are related by a boundary Legendre transform. It was noted in \cite{brown} that the Brown-York action 
is related to the Gibbons-Hawking action by a Legendre transform in the {\em canonical} variables. Not surprisingly, one can check that the Neumann action, \eqref{adm_action_neumann}, is {\em not} a Legendre transform of the Brown-York action, \eqref{Brown-York-action},  in terms of {\em canonical} variables. To see this, let us first note that \cite{brown}
\begin{equation}
\frac{\delta S_{BY}}{\delta(\sqrt{\sigma}\varepsilon)} = -N, \quad \frac{\delta S_{BY}}{\delta(\sqrt{\sigma}j_a)} = N^a, \quad \frac{\delta S_{BY}}{\delta \sigma_{ab}} = -\frac{N\sqrt{\sigma}s^{ab}}{2}
\end{equation}
The Neumann action, \eqref{adm_action_neumann}, can be now written as
\begin{equation}
S_N = S_{BY} - \int d^{D-1}x \left[ \frac{1}{2}\frac{\delta S_{BY}}{\delta(\sqrt{\sigma}\varepsilon)}(\sqrt{\sigma}\varepsilon) + \frac{\delta S_{BY}}{\delta(\sqrt{\sigma}j_a)}(\sqrt{\sigma}j_a) + \frac{\delta S_{BY}}{\delta \sigma_{ab}}\sigma_{ab}   \right]
\end{equation}
The factor of $1/2$ spoils the Legendre transform and is the reason why the variation of the Neumann action, \eqref{neu_act_var}, has mixed terms proportional to variations of $\sqrt{\sigma}\varepsilon$ as well as $N$. So the Neumann ensemble is best thought of a mixed ensemble where there is dependence both on $N$ and $\sqrt{\sigma}\varepsilon$ or equivalently, on temperature and energy density. 

This means that a naive Laplace transform of the Brown-York partition function (aka microcanonical density of states) of the form (say)
\begin{equation}
Z_N \neq \int D(\sqrt{\sigma}\varepsilon) D(\sqrt{\sigma}j_a \phi^a)D(\sigma_{ab})\nu[\varepsilon,j_a,\sigma_{ab}]\exp\left[ \int_{B}d^{D-2}\theta \sqrt{\sigma}\beta \left( \frac{\varepsilon}{2} - \omega j_a \phi^a + \frac{1}{2}\mu^{ab}\sigma_{ab} \right) \right]
\end{equation}
cannot work: this is because if we integrate the right hand side over $D(\sqrt{\sigma}\varepsilon)$ the right hand side will purely be a function of temperature and will not have any dependence on $\epsilon$, which cannot be the case for the Neumann partition function. To get to the true Neumann partition function, in principle one must identify the function of temperature and energy that characterizes the ensemble and set it to a constant in the path integral via a delta functional. For our purposes of showing the emergence of the correct Smarr formula and first law, fortunately these subtleties will not be necessary.


For our purposes, we merely have to determine the complex saddles of 
\bea Z_N=\sum_M\int D[H]\  \exp(i S_N)\label{NeumPartfn}
\eea
which when we demand smoothness, again force the correct periodicity of the time circle. Together with the addition of the horizon term (\ref{BH_dos_action}) to ensure the right boundary condition at the ``bolt'' and a regularization scheme in the asymptotic region, this will make our partition function well-defined and computable (in the saddle point approximation). The regularization scheme will depend on the asymptotics of the geometry: in this paper we will consider asymptotically flat situations and use a form of background subtraction, in a companion paper we will consider the asymptotically AdS case and develop a version of holographic renormalization.

Analogous to the Brown-York/Gibbons-Hawking cases, one can choose to think of the black hole in the co-rotating frame. 
But we will see that this is not strictly necessary (in any of the ensembles) when interpreted correctly, because the horizon piece in the action (\ref{BH_dos_action}) will get a contribution that automatically implements this as a sort of ``datum subtraction''. Also for stationary flat space (Kerr) black holes, we will find that the $\mu^{ab}\equiv\frac{N\sqrt{\sigma}s^{ab}}{2} $ fall off sufficiently fast that they don't contribute in the discussion of the Smarr formula.

This Neumann ensemble for gravity, we will use in the next section to discuss horizon thermodynamics. Before we proceed however, we make one comment. We will use notations like $\beta, M$ to denote quantities in the Neumann ensemble as well, even though they are \emph{not} the defining thermodynamical quantities in the Neumann ensemble. Both these objects are well-defined geometrically: $\beta$ is fixed by the periodicity of the time circle via smoothness of the ``bolt'' as we discussed. Also, since $N \rightarrow 1$ in asymptotically flat space, $\sqrt{\sigma} \varepsilon/ 2N$ (which is what one holds fixed in Neumann), $\rightarrow \sqrt{\sigma} \varepsilon/2$. The integral of this quantity over $d^{D-2} \theta$ is what we call $M/2$ (after suitable background subtraction), and this is a well-defined geometric quantity in any saddle as well. So we will express our Neumann thermodynamic relations in terms of them. This philosophy should be compared to the discussion of how the time periodicity is fixed in Section  IV (and thermodynamics in Section VI) of \cite{brown}: the ensemble one starts with there is microcanonical, but $\beta$ is a nonetheless useful quantity.

\section{Horizon Thermodynamics}

In this section, as outlined many times previously, we will compute the Dirichlet and Neumann actions both covariantly and canonically (after doing an appropriate background subtraction). Equating the covariant and canonical results to each other  will reproduce the generalized Smarr formula in both Dirichlet and Neumann.  We will do this for Schwarzschild black holes in all dimensions and the 4-dimensional Kerr black hole. In the original Gibbons-Hawking computation \cite{GHY}, the Smarr formula was obtained in the Dirichlet ensemble by setting the (covariantly obtained) free energy to $M-T S - \Omega J$, but it is well known that this last expression can also be obtained canonically. We repeat it here for convenient comparison with the Neumann case.

\subsection{Dirichlet}

\subsubsection{Schwarzschild in D dimensions}

The appropriate background subtracted Dirichlet action in $D$ dimensions in the covariant form is
\begin{equation}
S_D = \frac{1}{2\kappa}\int_{\cal M}d^Dx\sqrt{-g}R + \frac{1}{\kappa}\int_{\cal B}d^{D-1}x \sqrt{-\gamma}(\Theta - \Theta_0)\label{subDir}
\end{equation}
where $\Theta_0$ is the extrinsic curvature of the boundary embedded in Minkowski space. The thermodynamics of the black hole can be understood by evaluating the complex metric associated with \eqref{lorentzian_metric} on this action. The associated complex metric is given by the identification $N = -i \tilde{N}$ and has periodically identified time with periodicity $\beta$. We consider the boundary to be at $r=R_c$ which shall be pushed to infinity eventually.

For the Schwarzschild black hole, we have
\begin{eqnarray}
\Theta &=& \frac{\sqrt{1-\frac{2M}{R_{c}^{D-3}}}\left((D-2)R_{c}^{D-3} -(D-1)M \right)}{R_{c}^{D-2} - 2M R_c} \\ \nonumber
\Theta_0 &=& \frac{D-2}{R_c} \nonumber
\end{eqnarray}
Evaluating the action and taking $R_c \rightarrow \infty$ limit, we get
\begin{equation}
S_D = i \frac{S_{D-2}(2M)^{\frac{D-2}{D-3}}}{4(D-3)}
\end{equation}
Now this is related to the Dirichlet/grand canonical/Helmholtz free energy by the relation
\begin{equation}\label{D_free_energy}
-\beta F_D \equiv \log Z_D  \approx iS_D
\end{equation}
This gives 
\begin{equation}
F_D^{cov} = \frac{S_{D-2}M}{8\pi}=\frac{M_{ADM}}{D-2},
\end{equation}
where in the last step we have used the definition of ADM mass given in the Appendix.

Now, we consider the canonical computation. After the background subtraction, the ADM version of the action for Dirichlet is given by
\begin{eqnarray} \nonumber
S_D &=& \int_{\cal M}d^D x\left[ p^{ab}\dot{h}_{ab} - NH - N^a H_a \right] + \int_{\cal H}d^{D-1}x \sqrt{\sigma}\left(\frac{r^a\partial_a N}{\kappa} + \frac{2N^a r^b p_{ab}}{\sqrt{h}}  \right) \\ \nonumber &+& \int_{\cal B}d^{D-1}x \sqrt{\sigma}\left[ N(\varepsilon - \varepsilon_0) - N^a (j_a - j_{a\;0}) \right] 
\end{eqnarray}
For the Schwarzschild metric
\begin{eqnarray}
N^a = 0, \ \ k = \frac{(D-2)\sqrt{1-\frac{2M}{R_{c}^{D-3}}}}{R_c}, \ \ k_0 = \frac{D-2}{R_c}.
\end{eqnarray}
As we explained in the general case in the previous section, from the arguments of Brown and York, the horizon integral evaluated over the complex metric gives
\begin{equation}
S_{\cal H} = -i\frac{A}{4}
\end{equation}
The boundary integral on the other hand evaluates to
\begin{equation}
S_{\cal B}  = \frac{i}{4}\left( \frac{D-2}{D-3} \right)S_{D-2}(2M)^{\frac{D-2}{D-3}}
\end{equation}
Using \eqref{D_free_energy}, we get
\begin{equation}
F_D^{canon} = \frac{(D-2)S_{D-2}M}{8\pi} -\frac{A}{4\beta}
\end{equation}
which is exactly of the form (see Appendix)
\begin{equation}
F_D^{canon}= M_{ADM} - TS
\end{equation}
One can check that setting this equal to the covariant result $F_D^{cov}=F_D^{canon}$ is the correct Smarr formula 
\bea
\frac{D-3}{D-2}M_{ADM}=TS \label{Dsmarr}.
\eea

\subsubsection{Kerr in 3+1 dimensions}

The thermodynamics of Kerr black hole can be understood by evaluating the complex metric associated with \eqref{kerr_lorentzian_metric} on the Dirichlet action (\ref{subDir}). 
The asymptotic form of $\Theta$ and $\Theta_0$ is given by
\begin{eqnarray}
\Theta &=& \frac{2}{R_c} - \frac{M}{R_c^2} + O(1/R_c^3)\\ \nonumber
\Theta_0 &=& \frac{2}{R_c}+ O(1/R_c^3) \nonumber
\end{eqnarray}
where $R_c$ is the cut-off radius. 
The relevant complex metric associated to Kerr is given by the identification $N = -i \tilde{N}$, $N^{\phi} = -i \tilde{N}^{\phi}$ and has periodically identified time with periodicity $\beta$ which can be calculated by evaluating $r^a\partial_a N = 2\pi/\beta$ term in the ADM split action. This gives the periodicity 
\begin{equation}
\beta = \frac{4\pi r_H (r_H^{2}+a^2)}{(r_{H}^2 - a^2)}
\end{equation}
We consider the boundary to be at $r=R_c$ which shall be pushed to infinity eventually. Evaluating the action and taking $R_c \rightarrow \infty$ limit, we get

\begin{equation}
S_D = \frac{iM \beta}{2}
\end{equation}
In order to have a thermodynamic interpretation, the periodicity $\beta$ should be identified with the inverse temperature obtained by the Euclidean arguments. This gives, upon using (\ref{D_free_energy}):
\begin{equation}\label{cov_free_energy}
F_D^{cov}= \frac{1}{2}M
\end{equation}

Now we turn to the canonical calculation. The form of the background subtracted ADM action for the Dirichlet case was given in the Schwarzschild discussion. Restricted to 3+1 dimensions, it takes the form: 
\begin{eqnarray}
S_D &=& \int_{\cal M}d^4 x\left[ p^{ab}\dot{h}_{ab} - NH - N^a H_a \right] \\ \nonumber &+& \int_{\cal H}d^{3}x \sqrt{\sigma}\left(\frac{r^a\partial_a N}{\kappa} + \frac{2N^a r^b p_{ab}}{\sqrt{h}}  \right) + \int_{\cal B}d^{3}x \sqrt{\sigma}\left[ N(\varepsilon - \varepsilon_0) - N^a j_a \right] 
\end{eqnarray}
Since we are subtracting Minkowski space, $j_{a \ 0}=0$ for Kerr as well. The relevant non-trivial quantities can be computed to be
\begin{eqnarray}
\varepsilon - \varepsilon_0 &=& -\frac{2M}{\kappa}\frac{1}{R_c^2} + O(1/R_c^3) \\ \nonumber
j^{\phi} &=& \frac{6aM}{\kappa} \frac{1}{R_c^4} + O(1/R_c^5)
\end{eqnarray}
Thus we see that $N^{\phi}j_{\phi}$ term does not contribute at the boundary in the limit, $R_c \rightarrow \infty$. The horizon integral evaluated over the complex metric gives
\begin{equation}
S_{\cal H} = -i\frac{A}{4} -i\Omega_H aM \beta
\end{equation}
Note the extra piece, which vanishes in the co-rotating frame (but then it would have appeared as a boundary piece, so that the final Smarr formula remains intact).
The boundary integral evaluates to
\begin{equation}
S_{\cal B}  = iM\beta
\end{equation}
Using \eqref{D_free_energy} for the free energy, we get
\begin{equation}
F_D^{canon} = M - TS - \Omega_H J
\end{equation}
Equating this to the covariant result \eqref{cov_free_energy}, we reproduce the 4D Smarr formula
\begin{equation}
\frac{1}{2}M = TS + \Omega_H J
\end{equation}

\subsection{Neumann}

\subsubsection{Schwarzschild in D dimensions}

We will define the background subtracted Neumann action in covariant form by analogy with the Dirichlet case:
\begin{equation}
S_N = \frac{1}{2\kappa}\int_{\cal M}d^Dx \sqrt{-g} R + \frac{(4-D)}{2\kappa}\int_{\cal B}\sqrt{-\gamma}(\Theta - \Theta_0) \label{regulator}
\end{equation}
Evaluating the action for the complex metric, we get
\begin{equation}
S_N = -i \frac{(D-4)S_{D-2}(2M)^{\frac{D-2}{D-3}}}{8(D-3)}
\end{equation}
This is related to the Neumann free energy by
\begin{equation}\label{N_free_energy}
-\beta F_N \equiv \log Z_N  \approx iS_N
\end{equation}
which gives 
\begin{equation}
F_N^{cov} = -\frac{(D-4)S_{D-2}M}{16\pi}=-\frac{(D-4)}{2(D-2)}M_{ADM}
\end{equation}

Now, the background subtracted Neumann action in ADM variables is given
\begin{eqnarray}\nonumber
S_N &=& \int_{\cal M}d^D x\left[ p^{ab}\dot{h}_{ab} - NH - N^a H_a \right] + \int_{\cal H}d^{D-1}x \sqrt{\sigma}\left(\frac{r^a\partial_a N}{\kappa} + \frac{2N^a r^b p_{ab}}{\sqrt{h}}  \right) \\ \nonumber &+& \int_{\cal B}d^{D-1}x\sqrt{\sigma}\left[ \frac{N}{2}(\varepsilon-\varepsilon_0) - N^a (j_a - j_{a\;0}) + \frac{N}{2}(s^{ab}-s_{0}^{ab})\sigma_{ab} \right]
\end{eqnarray}
Evaluating the action on the complex metric, the horizon integral gives a contribution
\begin{equation}
S_{\cal H} = -i\frac{A}{4}
\end{equation}
The boundary integral evaluates to
\begin{equation}
S_{\cal B}  = i\left( \frac{D-2}{D-3} \right)S_{D-2}(2M)^{\frac{D-2}{D-3}}
\end{equation}
Using \eqref{N_free_energy} the free energy takes the form
\begin{equation}
F_N = \frac{(D-2)}{16\pi}S_{D-2}M - \frac{A}{4 \beta}
\end{equation}
which is of the form
\begin{equation}
F_N^{canon} = \frac{1}{2}M_{ADM} - TS
\end{equation}
Equating $F^{canon}=F^{cov}$ again leads to the correct Smarr formula (\ref{Dsmarr}) as it did in the Dirichlet ensemble.

\subsubsection{Kerr in 3+1 Dimensions}

The Neumann action in covariant form in 4D is given by
\begin{equation}
S_N = \frac{1}{2\kappa}\int_{\cal M}d^4x \sqrt{-g}R
\end{equation}
Since Kerr solution is Ricci flat, the on-shell action for the complex metric vanishes. Furthermore, this leads to zero Neumann free energy
\bea
F_N^{cov}=0.
\eea

The Neumann action in terms of the ADM variables in 4D is given
\begin{eqnarray}\nonumber
S_N &=& \int_{\cal M}d^4 x\left[ p^{ab}\dot{h}_{ab} - NH - N^a H_a \right] + \int_{\cal H}d^{3}x \sqrt{\sigma}\left(\frac{r^a\partial_a N}{\kappa} + \frac{2N^a r^b p_{ab}}{\sqrt{h}}  \right) \\ \nonumber &+& \int_{\cal B}d^{3}x\sqrt{\sigma}\left[ \frac{N}{2}(\varepsilon-\varepsilon_0) - N^a j_a + \frac{N}{2}(s^{ab}-s_{0}^{ab})\sigma_{ab} \right]
\end{eqnarray}
Evaluating the action on the complex metric, the horizon integral gives a contribution
\begin{equation}
S_{\cal H} = -i\frac{A}{4} - i\Omega_H a M \beta
\end{equation}
On the boundary, we have
\begin{equation}
s^{ab}-s_{0}^{ab} = -\frac{M^2}{2\kappa}\left( \begin{array}{cc}
1 & 0 \\
0 & \frac{1}{\sin^2 \theta}
\end{array} \right)\frac{1}{R_c^5} + O(1/R_c^6)
\end{equation}
and 
\begin{equation}
\sigma_{ab} = \left( \begin{array}{cc}
\rho^2 & 0 \\
0 & \frac{\sin^2 \theta}{\rho^2}\left( (r^2 + a^2)^2 - a^2 \Delta \sin^2 \theta \right)
\end{array} \right)
\end{equation}
Only the first term in the boundary integral contributes, while the other terms fall-off rapidly when the boundary is pushed to infinity.
\begin{equation}
S_{\cal B}  = i\frac{M}{2}\beta
\end{equation}
The free energy is computed via (\ref{N_free_energy}) to be
\begin{equation}
F_N^{canon} = \frac{1}{2}M - TS- \Omega_H J.
\end{equation}
Setting this equal to $F_N^{cov}=0$ obtained above shows that Neumann ensemble also correctly reproduces the Smarr formula.

\subsection{Cosmological Horizons}

A similar approach can also be applied to de-Sitter geometry, whose metric in the static coordinates is given by $ ds^2 = -(1-\frac{r^2}{\alpha^2})dt^2 + \frac{dr^2}{(1-\frac{r^2}{\alpha^2} )} + r^2d\Omega^{2}_{(D-2)} $. The crucial difference between the de-Sitter and the black hole geometries considered previously is that the relevant complexified section of the geometry does not have any boundaries. Therefore all the contributions comes from the horizon term\footnote{We believe the action for de Sitter mentioned in eqn (3.15) of \cite{GHY} has a numerical factor of 4 missing. Our result matches the one quoted in \cite{Hartmann, BoussoDeSitterReview}.}. As in the black hole case, regularity of the complexified de-Sitter at the horizon at $r=\alpha \equiv \sqrt{3/|\Lambda|}$ (where $\Lambda$ is the cosmological constant), fixes the periodicity of the time coordinate to be $P=2\pi\alpha$. Finally, the on-shell action evaluates to $S_{dS}= -\frac{2\pi i}{\kappa} \int d^{D-2}\theta\;\sqrt{\sigma} = -3\pi i\Lambda^{-1}$. Thus we find that the action used to compute black hole density of states is also suitable for computing the entropy of de-Sitter space. This is because in both the cases the crucial argument is the periodicity of the time circle which relies on the fact that the horizon is a bifurcate Killing horizon which is true in both cases.

\subsection{First Law}

We will conclude this section by deriving the first law from Neumann path integral around its saddles. We will work by analogy with the microcanonical discussion in \cite{brown}. The variations of the Neumann action are given by \eqref{neu_act_var}, which when restricted around the complex saddles takes the form
\begin{eqnarray}
\delta S_N = -i\int d^{D-1}y \left[ \delta \tilde{N}\left(\frac{\sqrt{\sigma}\varepsilon}{2}\right) -  \tilde{N} \delta \left(\frac{\sqrt{\sigma}\varepsilon}{2} \right) - \delta \tilde{N}^a \left(\sqrt{\sigma}j_a \right) - \delta \left( \frac{\tilde{N}\sqrt{\sigma}s^{ab}}{2} \right)\sigma_{ab}  \right]
\end{eqnarray}
Using the relation between free energy and on-shell action and noting that on the solutions we consider, asymptotically at the boundary, the conditions
\begin{equation}
\int dt \tilde{N} = \beta, \quad \int dt \tilde{N}^{\phi} = \beta \Omega
\end{equation}
hold\footnote{$N \rightarrow 1$ at the boundary, and so we treat $\beta$ as the periodicity of the time circle, which is fixed by the smoothness of the bolt at the horizon. Note also that our conventions for extrinsic curvature are opposite to those in \cite{brown}, so the charges are defined with an extra negative sign. See also the discussion at the end of Section 4, for the meaning assigned to $\beta$ in the Neumann ensemble.}, we get the functional form
\begin{eqnarray}
\delta(iS_N) &=& \delta(-\beta F_N) \\ \nonumber &=& -\int_{B}d^{D-2}\theta \left[ - \delta \beta \left(\frac{\sqrt{\sigma}\varepsilon}{2} \right) + \beta \delta \left(\frac{\sqrt{\sigma}\varepsilon}{2} \right) + \delta(\beta \omega)(\sqrt{\sigma}j_a \phi^a) + \delta \left( \frac{\beta\sqrt{\sigma}s^{ab}}{2} \right)\sigma_{ab} \right]
\end{eqnarray}
The above expression is of the form (see \cite{brown} for the definition of pressure $p$),
\begin{equation}
d(-\beta F_N) = -d\beta \frac{E}{2} + \frac{\beta}{2}dE - d(\beta \Omega)J - d(\beta p)V
\end{equation}
Using the Hamiltonian form for the Neumann free energy that we obtained in the previous sections,
\begin{equation}
F_N = \frac{E}{2}- TS - \Omega J + pV
\end{equation}
we get the familiar form of the first law
\begin{equation}
TdS = dE - \Omega dJ + pdV.
\end{equation}

\section{Comments and Speculations}

We have discussed a path integral for gravity with new boundary conditions and seen that it leads to various reasonable conclusions. 

In particular, we have seen that thermodynamics of horizons emerges naturally, just as it did in the Gibbons-Hawking and Brown-York ensembles. The essential reason for this is the fact that the boundary terms merely change the ensemble, the basic thermal quantities are the temperature and the entropy, both of which are found at the horizon. This leads us naturally to the idea that any reasonable boundary term for gravity should indeed lead to consistent thermodynamics. The results for Gibbons-Hawking, Brown-York and Neumann are merely special instantiations of this general principle.

It will also be interesting to extend this work to higher dimensions where there are other kinds of horizons \cite{5d}. We expect that the discussion here should apply there as well, after minor modifications.

\section*{Acknowledgments}

CK thanks Shahin Sheikh-Jabbari for correspondence. We are grateful to Sumanta Chakraborty, Bob McNees, Inyong Park and Dmitri Vassilevich for comments and/or pointers to references after \cite{KR} appeared. We thank Pallab Basu, Justin David, Ghanashyam Date, Jarah Evslin, Sudipta Sarkar, Shahin Sheikh-Jabbari, Kostas Skenderis and Sergey Solodukhin for questions, comments and/or discussions. CK thanks the organizers and audiences at conferences at Bangalore, Phitsanulok and Yerevan where some of this material was presented, for stimulating discussions and hospitality.

\appendix

\section{Schwarzschild-Tangherlini Conventions}
The Schwarzschild metric in $D$ dimensions is given by

\begin{equation}\label{lorentzian_metric}
ds^2 = -f(r)dt^2 + \frac{dr^2}{f(r)} + r^2 d\Omega^{2}_{(D-2)}, \qquad f(r) = 1-\frac{2M}{r^{D-3}}
\end{equation}
where $M$ is the black hole mass parameter, related to the ADM mass of the black hole as \cite{Peet}

\begin{equation}
M_{ADM} = \frac{(D-2)S_{D-2}M}{8\pi}
\end{equation}
where $S_{n}$ is the area of the unit $n$-sphere and obeys the relation
$S_{n-1} = \frac{2\pi^{n/2}}{\Gamma \left(\frac{n}{2}\right)}$.
Now \eqref{lorentzian_metric} is a vacuum solution of Einstein's equation with $\Lambda =0$ and satisfy $R_{\mu \nu}=0$. The horizon is at

\begin{equation}
r_H = (2M)^{\frac{1}{D-3}}
\end{equation}
and the inverse temperature is given by

\begin{equation}\label{schtemp}
\beta = \frac{4\pi}{f'(r_H)} = \frac{4\pi (2M)^{\frac{1}{D-3}}}{(D-3)}
\end{equation}

\section{Kerr Conventions}
The $D=4$ Kerr metric in Boyer-Lindquist coordinates is given by
\begin{eqnarray}\label{kerr_lorentzian_metric}
ds^2 &=& \rho^2\left( \frac{dr^2}{\Delta} + d\theta^2 \right) + \frac{\sin^2 \theta}{\rho^2}\left( adt - (r^2 + a^2)d\phi \right)^2 \\ \nonumber  &-& \frac{\Delta}{\rho^2}\left( dt - a \sin^2 \theta d\phi \right)^2
\end{eqnarray}
where $a = \frac{J}{M}$ is the angular momentum parameter and $\rho^2 = r^2 + a^2 \cos^2 \theta$ and  $\Delta = r^2 + a^2 -2Mr$.
This is an asymptotically flat solution to the vacuum Einstein's equation with zero cosmological constant and describes the geometry of a rotating black hole. The horizon is located at the largest positive root of $\Delta(r_H)=0$ and is given by
$r_H = M + \sqrt{M^2 -a^2}$ and the angular velocity at the horizon is given by 
\begin{equation}
\Omega_H = \frac{a}{r_H^2 + a^2}
\end{equation}
Comparing the metric \eqref{kerr_lorentzian_metric} with the ADM split metric, the lapse and shift functions can be extracted as
\begin{eqnarray}
N &=& \sqrt{\frac{\rho^2 \Delta}{(r^2 + a^2)^2 - a^2 \Delta \sin^2 \theta}} \\ \nonumber
N^{\phi} &=& -\frac{2aMr}{(r^2 + a^2)^2 - a^2 \Delta \sin^2 \theta} \nonumber
\end{eqnarray}

\end{document}